\documentclass[useAMS,usenatbib]{mn2e}
\usepackage{graphicx}

\title[W Vir stars with K2]{First observations of W Virginis stars with K2: detection of period doubling}
\author[E. Plachy et al.]{E. Plachy$^{1}$\thanks{E-mail:
eplachy@konkoly.hu}, L. Moln\'ar$^{1}$, M. I. Jurkovic$^{1,2}$, R. Smolec$^{3}$, P. A. Moskalik$^{3}$, A. P\'al$^{1,4}$, \and L. Szabados$^{1}$, R. Szab\'o$^{1}$ \\\\
$^{1}$Konkoly Observatory, MTA CSFK, Konkoly Thege Mikl\'os \'ut 15-17, H-1121 Budapest, Hungary\\
$^{2}$Astronomical Observatory of Belgrade, Volgina 7, 11 060 Belgrade, Serbia \\
$^3$Copernicus Astronomical Center, Polish Academy of Sciences, ul.\ Bartycka 18, 00-716 Warszawa, Poland\\
$^4$E\"otv\"os Lor\'and University, P\'azm\'any P\'eter s\'et\'any 1/A, H-1117 Budapest, Hungary\\
}
\begin{document}

\date{Accepted. Received; in original form }

\pagerange{\pageref{firstpage}--\pageref{lastpage}} \pubyear{2002}

\maketitle

\label{firstpage}

\begin{abstract}
We present the first analysis of W Vir stars observed by the \textit{Kepler} space telescope in the K2 mission. Clear cycle-to-cycle variation were detected in the light curves of KT~Sco and the globular cluster member M80-V1. While the variations in the former star seems to be irregular on the short time scale of the K2 data, the latter appears to experience period doubling in its pulsation. Ground-based colour data confirmed that both stars are W Vir-type pulsators, while a comparison with historical photometric time-series data revealed drastic period changes in both stars. For comparison we reexamine ground-based observations of W~Vir, the prototype of the class, and conclude that it shows period doubling instead of mode beating. These results support the notion that nonlinear dynamics plays an important role in the pulsation of W Virginis-type stars.

\end{abstract}


\section{Introduction}

W Virginis stars are population II, low mass stars that likely undergo blueward loops away from the asymptotic giant branch due to helium shell flashes \citep{gingold, wallerstein}. They represent the intermediate period ($P$=$\sim$4--20~d) subclass of the Type II Cepheid variables, between the BL~Her and RV~Tau stars. We note, however, that the boundaries between the subclasses are not well-defined \citep{welch}. 
Most Type II Cepheids pulsate in the radial fundamental mode, showing more irregularities towards the longer periods. In the light curves of RV Tau stars irregularities are observed in addition to the period doubling (PD), the alternation of lower- and higher-amplitude cycles that characterize this subtype. But PD is not universally observed in the $P>20$~d period range, and appears in a few stars with shorter periods too \citep{wils,ogle2011}. The possibility of PD was mentioned but rejected by \citet{templeton2007} in their study of the prototype star W Vir. PD can be also found among BL~Her stars \citep{ogle2011, bl}, and space-based photometry revealed that RR Lyrae stars may show it in their light curves as well \citep{pd_rrl, smolec2015}. 



The most promising explanation of PD involves nonlinear dynamics, where period doubling bifurcation naturally occurs at certain parameter combinations of the system. The early hydrodynamical models \citep{model1, model2} already predicted nonlinear phenomena in W Vir stars: PD and even chaos was presented. The bifurcation, e.g., the destabilisation of the single-period limit cycle was caused by half-integer resonances between the fundamental mode and the first overtone (3:2 resonance) or the second overtone (5:2~resonance). Recent theoretical calculations suggest that 7:2 resonance with the third and 9:2 resonance with the fourth overtone are also plausible but only for narrower parameter ranges \citep{model_smolec}. Two PD domains were identified: one contains the BL~Her type and the short-period W Vir stars in the period range of $\sim$2-6.5~d, while the other appears at periods longer than $P>9.5$~d.

Large ground- and space-based surveys have potential to map these domains, although the former are limited by accuracy, and the latter by small sample size per mission so far. The only Type II Cepheid observed from space, the BL~Her subtype \textit{CoRoT} 0659466327 is outside the theoretical PD domains with its 1.542353 d period, and indeed does not show cycle-to-cycle variation \citep{CoRoT}. Fortunately, the K2 mission will build up a large collection of variable stars, including hundreds of Type II Cepheid stars in the coming years \citep{plachy2016}.

In this paper we report the analysis of the first space observations of W Vir-type stars: KT Sco, a field star, and M80-V1, a member of the globular cluster M80. KT~Sco ($\alpha_{2000.0}=16^{\rm h}$11$^{\rm m}$14\fs90, $\delta_{2000.0}$ = $-16$\degr 51\arcmin 40\farcs 2, $V$ = 12.48 mag) first appeared as an unclassified variable star in the list of \citet{guthnick1934}, but was apparently neglected afterwards. It was finally classified as a W Vir star with the advent of automated variability surveys \citep{asas,hoffman2009}.  The variation of M80-V1 ($\alpha_{2000.0}=16^{\rm h}$16$^{\rm m}$52\fs60, $\delta_{2000.0}$ = $-22$\degr 57\arcmin 15\farcs 9, $V$ = 13.42 mag) was discovered by \citet{bailey1902} in his extensive work on globular cluster variables. The star was later identified as a long-period Cepheid pulsating in the fundamental mode and it was regularly monitored during the 20th century \citep{sawyer1942,wehlau1984,wehlau1990,jim,kopacki2003}. We note that M80 belongs to the old population of globular clusters, with an age of $12.4\pm1.1$ Gyr and an average metallicity in the range of [Fe/H] = --1.47 and --1.75 \citep{salaris,marin-franch}.

Details of the observations and data reduction are given in Section 2. Section~3 presents the light curve analysis, the decadal period changes, and colour-magnitude diagrams for both stars. In Section 4 re-analyse the star W Vir, and discuss the PD interpretation of its pulsation behaviour. Finally we summarise our results in Section 5.

\section{Observations and data reduction}

\begin{figure}
\includegraphics[width=1.0\columnwidth]{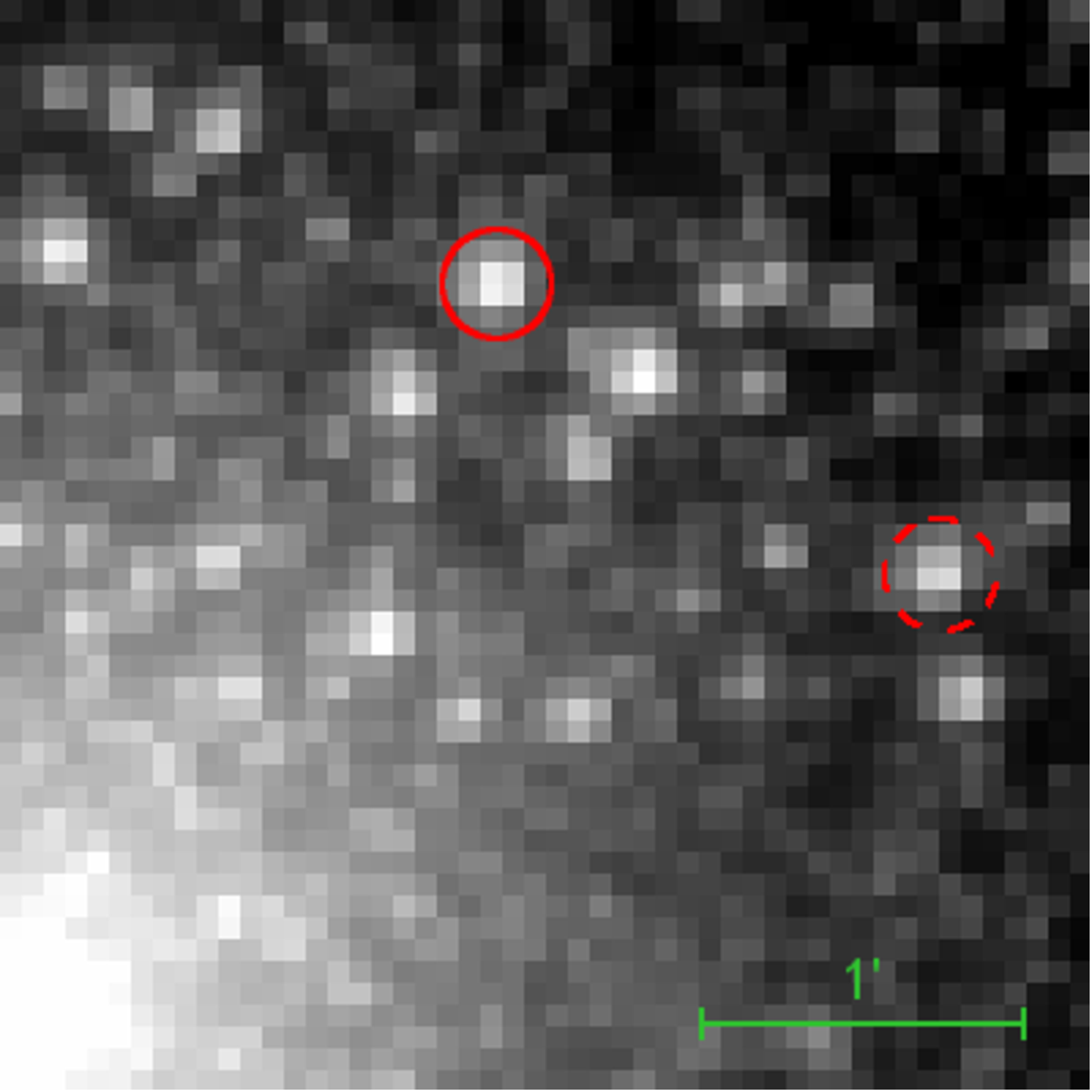}
\caption{The K2 stamp EPIC~200004389 that contains M80-V1 (EPIC~204281168). Red circle marks the aperture we used for the photometry. The dashed circle marks EPIC 204287127 that we used for comparison. }
\label{stamp}
\end{figure}

\begin{figure}
\includegraphics[width=1.0\columnwidth]{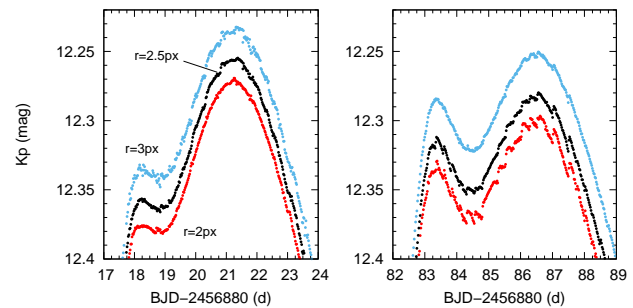}
\caption{Details of the photometry of M80-V1 with different apertures. Left and right panels show the first and last pulsation maxima. The small jumps are caused by the attitude changes of \textit{Kepler}. The $r=2$ px aperture is strongly affected by the jumps towards the end. The $r=3$ px aperture is more affected at the beginning of the campaign.}
\label{stab}
\end{figure}

\begin{figure*}
\includegraphics[width=1.0\textwidth]{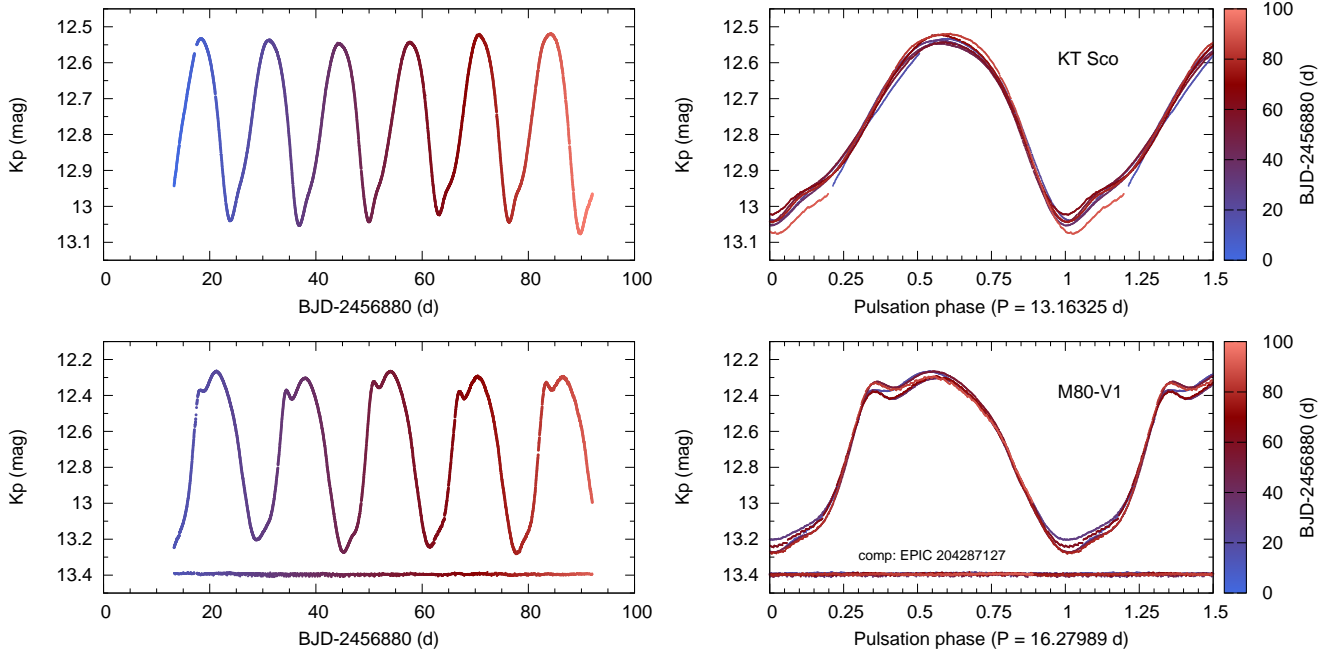}
\caption{The final K2 light curve (left) and folded phase curve (right) of KT~Sco, M80-V1, and EPIC 204287127, another bright star from M80 for comparison. Uncertainties are smaller than the symbols. The colour coding follows the temporal evolution of the light curve. The cycle-to-cycle variations are apparent in both cases, especially if we compare M80-V1 to another star from the globular cluster. }
\label{ktsco_lc}
\end{figure*}

Both stars were observed with the \textit{Kepler} space telescope during Campaign 2 of the K2 mission which lasted for 78.8 days, between August 23 and November 10, 2014. KT~Sco (EPIC 205546706) was observed as a regular target, while M80-V1 (EPIC 204281168, GSC 06793--01005) was included in the superstamp covering the globular cluster M80. Data were collected in long cadence mode (29.4 min sampling) in both cases. 

Photometry was obtained as follows. In the case of KT~Sco, we used the \textsc{pyke} software to create a custom target pixel mask for the star that contains the movements of the PSF caused by the drift of the telescope \citep{pyke}. Since the spacecraft has only two working reaction wheels, the radiation pressure from the Sun causes it to rotate around the third axis \citep{howell2014}. These movements are controlled with the onboard thrusters in every 6 h, but the images of the stars traverse between different pixels that leads to systematics in the flux measurements. Beside these movements, the pointing of \textit{Kepler} was adjusted by $\sim$2 pixels after the 50th cadence of the campaign that we also accommodated in the pixel mask. The variations of KT Sco were slow enough to separate the astrophysical signal from the characteristic jumps caused by the correction manoeuvres. We applied the self-flat-fielding (SFF) task of \textsc{pyke} to the residual light curve to remove the flux variations correlated with the motions of the centroid of the star, and then restored the assumed astrophysical component. 

In the case of M80-V1 the dense stellar field required a tighter mask with \textsc{pyke} to avoid contamination, but this in turn prevented us from covering the movements of the field. So we chose to apply a traditional aperture photometry using the \textsc{fitsh} software \citep{pal2012}. \textsc{Fitsh} is a fully open source astronomical data reduction and analysis software package, containing standalone binaries that are utilised via UNIX shell scripts. It has been used to extract the K2 observations of very faint RR Lyrae stars within the nearby dwarf galaxy Leo IV \citep{molnar2015}. 
 
M80 was covered with four 50$\times$50 px stamps. First we registered the frames of the stamp containing M80-V1 to the same reference system. Then we extracted light curves with different apertures between $r = 2-3$ px. The sizes of the stellar PSFs increased towards the end of the campaign, likely due to a shifting focus caused by the changing thermal environment of the telescope. None of the light curves were completely free from the small sawtooth-like jumps caused by the attitude control manoeuvres, but as Fig.~\ref{stab} shows, their appearance within the light curve changed with the size of aperture. We selected the light curve created with the aperture $r = 2.5$ px, and a sky aperture with an inner radius of 4 px and a width of 2 px for further analysis. One drawback of aperture photometry in dense fields is that other stars could easily contaminate the stellar and sky apertures. Luckily, M80-V1 is both luminous and lies at the outskirts of the globular cluster, at about 4.3 $R_h$, where $R_h = 37\arcsec$ is the half-light radius of the cluster (\citet{harris1996}, 2010 edition). With the selected parameters, the strongest contamination we identified is a $\sim$0.05 per cent leakage from M80-V1 itself into the sky aperture. 

Although the selected aperture is quite narrow, we found the obtained photometry stable enough for our analysis. The Leo IV study used even tighter apertures successfully, although for much fainter targets \citep{molnar2015}. We compared the light curve of M80-V1 to the light curves of other bright and relatively isolated stars in the field. One such example, EPIC 204287127 (GSC 06793-00817) is marked with a dashed circle in Fig.~\ref{stamp}. Its light curve shows small, $\sim0.01$ mag variations on 4--8 d long timescales that might be intrinsic to the star, and exceeds the scatter of the data points. These variations are significantly slower than the instrumental effects caused by the attitude control manoeuvres. Compared to the level of variations observed in M80-V1 (lower panel of Fig.~\ref{ktsco_lc}.), this latter light curve is essentially flat, indicating that instrumental effects are negligible in the case of our analysis. 

For further comparison, we applied the same differential-image photometric method that was used by \citet{molnar2015} to M80-V1, but the results did not exceed the quality of the light curve obtained by simple aperture photometry, so we decided to use the latter. The final photometric data are listed in Tables \ref{ktscodata} and \ref{m80data}. The light curves and the folded phase curves of the two stars are shown in Figure \ref{ktsco_lc}.

\begin{table}
\caption{Sample of the photometric data of KT Sco. The full table is accessible in the online version of the paper.}
\begin{center}
\begin{tabular}{ccc}
\hline
\hline
BJD (d)& \textit{Kp} mag  & $\Delta$\textit{Kp} (mag)  \\
\hline
2456893.283870 & 12.942613 & 0.000108\\
2456893.304302 & 12.938681 & 0.000108\\
2456893.324734 & 12.935734 & 0.000108\\
2456893.345166 & 12.934602 & 0.000108\\
2456893.365598 & 12.932009 & 0.000107\\
\multicolumn{3}{l}{\dots} \\ 
\hline
\end{tabular}  
\end{center}
\label{ktscodata}
\end{table}   

\begin{table}
\caption{Sample of the photometric data of M80-V1. The full table is accessible in the online version of the paper.}
\begin{center}
\begin{tabular}{ccc}
\hline
\hline
BJD (d)& \textit{Kp} mag  & $\Delta$\textit{Kp} (mag)  \\
\hline
2456893.284071 & 13.22563 & 0.00021\\
2456893.304503 & 13.22291 & 0.00021\\
2456893.324935 & 13.22105 & 0.00021\\
2456893.345367 & 13.21918 & 0.00021\\
2456893.365800 & 13.21909 & 0.00021\\
\multicolumn{3}{l}{\dots} \\ 
\hline
\end{tabular}  
\end{center}
\label{m80data}
\end{table} 

\section{Analysis}
\subsection{K2 light curves}

Thanks to the quality of K2 measurements, irregularities can be easily detected by visual inspection of the light curves. The brightness levels of the extrema, the amplitudes, and the shape of the consecutive pulsation cycles are all diverse in both stars. The bump/hump features that appear after minimum light and before maximum brightness also change slightly from one cycle to another.  These variations are even more prominent in the folded light curves (right panels of Figure \ref{ktsco_lc}). However, the details of the variations are different for the two stars. The light curve of KT Sco shows a non-repetitive nature but with only six pulsation cycles covered, the origin of this irregularity remains unclear. The possible causes can be period change, amplitude fluctuations, modulation or presence of an additional mode (or modes). M80-V1 shows an alternation of high- and low-amplitude cycles that resembles PD, although the repetition is not exact.

We performed a standard Fourier analysis using the Period04 software \citep{period04}. The Fourier solution for KT~Sco and M80-V1 is presented in Table \ref{freqtable}. We identified the $f_0=0.07597(3)$ c/d pulsation frequency and its four harmonics in the light curve of KT Sco. The pulsation frequency of M80-V1 was found to be $f_0=0.06142(3)$ c/d. In this star eight harmonics and two half-integer subharmonics ($0.5f_0$ and $1.5f_0$) were detected in the Fourier spectrum. The latter peaks correspond to the alternation of low- and high-amplitude cycles, the characteristic signature of PD. In this analysis we set the frequency values to the exact multiplets of the main peak. If we prewhiten the subharmonics with independent frequencies, the peaks are at ratios 0.486$f_0$ and 1.526$f_0$: the differences from the exact half-integers are well within the limited frequency resolution of the data. 

\begin{figure}
\includegraphics[width=1.0\columnwidth]{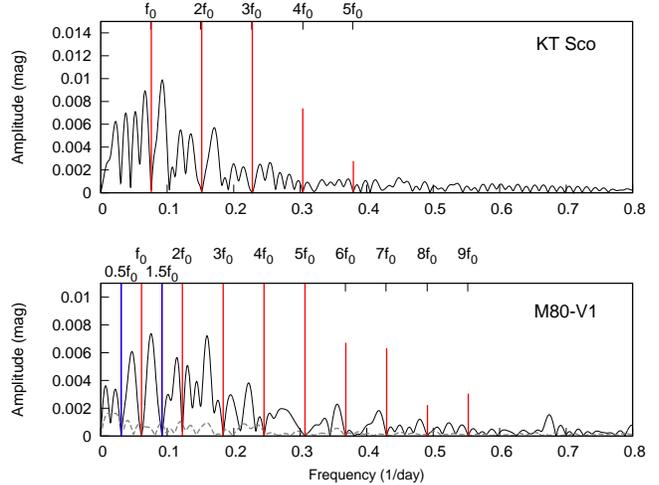}
\caption{Residual Fourier spectra of the two stars. Positions and amplitudes of the $nf_0$ harmonics are indicated with red lines, the half-integer peaks with blue ones. In the case of M80-V1, the grey dashed line shows the Fourier spectrum of the comparison target EPIC 204287127.}
\label{res}
\end{figure}

For the above fit we selected peaks with signal-to-noise ratios exceeding 3 for our frequency solution. The residual spectra still contain unresolved signals with significant power in the low-frequency region in both stars as it is displayed in Figure \ref{res}. They correspond to the temporal variations of the cycles we observed in the light curves, and could also indicate the possible presence of additional modes. For example, non-resonant beating between the fundamental mode and the fourth overtone was found in W Vir models recently with expected period ratios around ($P_4$/$P_0$) $\sim$0.3 \citep{model_smolec}. However we did not detect excess signals around the expected position of the fourth overtone in either star. 

The instrumental peaks that appear at $f_{\rm corr} \approx 4.05$ c/d and its harmonics are clearly separated from in the Fourier spectra. The amplitude of the $f_{\rm corr}$ peak is 0.27 mmag for KT~Sco, and 1.16 mmag for M80-V1: the large difference can be traced back to the SFF correction we applied to the data of KT Sco.


\begin{table}
\caption{Fourier parameters of KT Sco and M80-V1.}
\begin{center}
\begin{tabular}{clll}
\hline
\hline
Frequency& Frequency  & Amplitude  & \multicolumn{1}{c}{$\phi$}  \\
  ID &\multicolumn{1}{c}{(d$^{-1}$)} & \multicolumn{1}{c}{(mag)} & \multicolumn{1}{c}{(rad/2$\pi$)} \\
\hline
\noalign {\smallskip} 
\multicolumn{4}{c}{KT Sco} \\
\noalign {\smallskip} 

$f_0$ & 0.075973(10) & 0.2498(4) & 0.5075(3)\\ 
$2f_0$ & 0.15195(10) &  0.0299(4) & 0.958(2)\\  
$3f_0$ & 0.22792(15) & 0.0213(3) & 0.361(3)\\  
$4f_0$ & 0.3039(3) & 0.0074(4) & 0.711(10)\\ 
$5f_0$ & 0.3799(7) & 0.0028(10) & 0.943(15)\\ 
\hline
\multicolumn{4}{c}{M80-V1} \\

\noalign {\smallskip} 

$f_0$ &0.061424(4) &0.4932(3)& 0.75356(9)\\ 
$2f_0$ &0.12285(3)&0.0717(3)&0.2270(5)\\ 
$3f_0$ &0.18427(5)&0.0374(3)&0.2990(11)\\  
$4f_0$ &0.24569(5)&0.0403(3)&0.5972(10)\\ 
$5f_0$ &0.30712(9)&0.0216(3)&0.822(2)\\
$6f_0$ &0.3685(3)&0.0067(3)&0.041(6)\\
$7f_0$ &0.4230(3)&0.0063(3)&0.150(6)\\
$8f_0$ &0.4913(8)&0.0022(3)&0.35(2)\\
$9f_0$ &0.5528(6)&0.0031(3)&0.547(13)\\
$0.5f_0$ &0.03071(12)&0.0154(3)&0.613(3)\\
$1.5f_0$ &0.09213(12)&0.0150(3)&0.939(3)\\

\hline
\noalign{\smallskip}
\end{tabular}  
\end{center}
 \normalsize
 \label{freqtable}
\end{table}   
    
\subsection{Period changes}


To examine the long term variability of the stars, we collected observations from the literature. KT Sco was observed by a number of sky surveys starting from 1999. We used the measurements from the archives of the All-Sky Automated Survey (ASAS), the Northern Sky Variability Survey (NSVS), the Optical Monitoring Camera on board the INTEGRAL satellite (IOMC), and the Catalina Sky Survey (CSS)  (\citealt{asas,nsvs,omc,css}, respectively). 

In the case of M80-V1 we found historical data: photographic observations collected between 1939 and 1987 were published by \citet{wehlau1990}. Most of these were taken in Johnson $B$ band, but some $V$-band data were also obtained towards the end. The star was observed by the SuperWASP survey between 2006 and 2008 as well \citep{butters2010,paunzen2014}. The times of observations were converted to Barycentric Julian Date in all cases \citep{hjd}.

The exact shape of Type II Cepheid light curves, just as light curves of any other large-amplitude pulsators, depends on the photometric band in which it is observed. This may lead to a phase difference between the frequency fits of data obtained in different passbands. We had to take this effect into account in order to compare the older observations with the K2 data. \textit{Kepler} uses a white light-like wide passband, and the SuperWASP survey observes in white light, without any filters. As the spectral energy distribution of Type II Cepheids peak around the $R$ band, we approximated the phase difference between the photographic data with the $\Delta \phi_{B-R} = \phi_B-\phi_R$ and $\Delta \phi_{V-R} = \phi_V-\phi_R$ values. We used the extensive Johnson $BVRI$ observations of W Vir itself obtained by \citet{templeton2007} to  estimate the phase differences. The derived values are $\Delta \phi_{B-R} = 0.055$ and $\Delta \phi_{V-R} = 0.035$, in ${\rm rad}/2\pi$ units. Since the exact values might differ for M80-V1 and KT Sco slightly, we increased the uncertainties of the converted phase values with 0.01. 

\begin{figure}
\includegraphics[width=1.0\columnwidth]{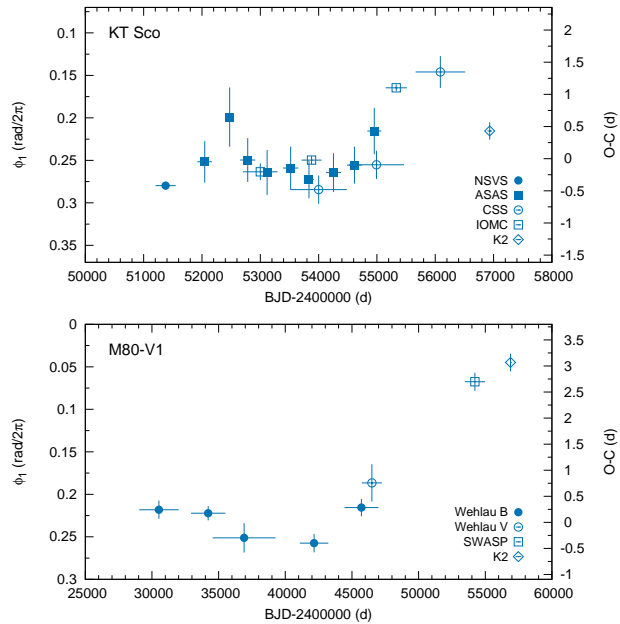}
\caption{Phase variation and O--C diagrams of both stars. Note that the timespans are different. The reference periods are 13.214985 d for KT Sco, based on the ASAS light curve, and 16.304545 d for M80-V1, based on the $B$-band data of \citet{wehlau1990}.  }
\label{o-c}
\end{figure}

Both phase variation (O--C) diagrams in Figure \ref{o-c} reveal considerable changes in the pulsation periods. KT~Sco experienced excursions towards longer periods but the K2 data indicate that its period is shortening again. The data hints at a possible periodic behaviour over a decadal time scale, but further observations will be needed to confirm it. The phase variation of M80-V1 suggests that it also experienced a significant period change during the 1970s (sometime between JD 2441000 and 2444000), and it has been pulsating with a longer period since then. These findings agree with previous studies: large and abrupt changes can often be observed in the periods of W Vir-type stars \citep{percy2000,rabidoux}. 

We also investigated if any amplitude change during the past decades could be detected. We found that the amplitudes remained constant for both stars within the uncertainties of the measurements and transformations between photometric bands.

\subsection{Colour-magnitude diagrams}
\begin{figure}
\includegraphics[width=1.0\columnwidth]{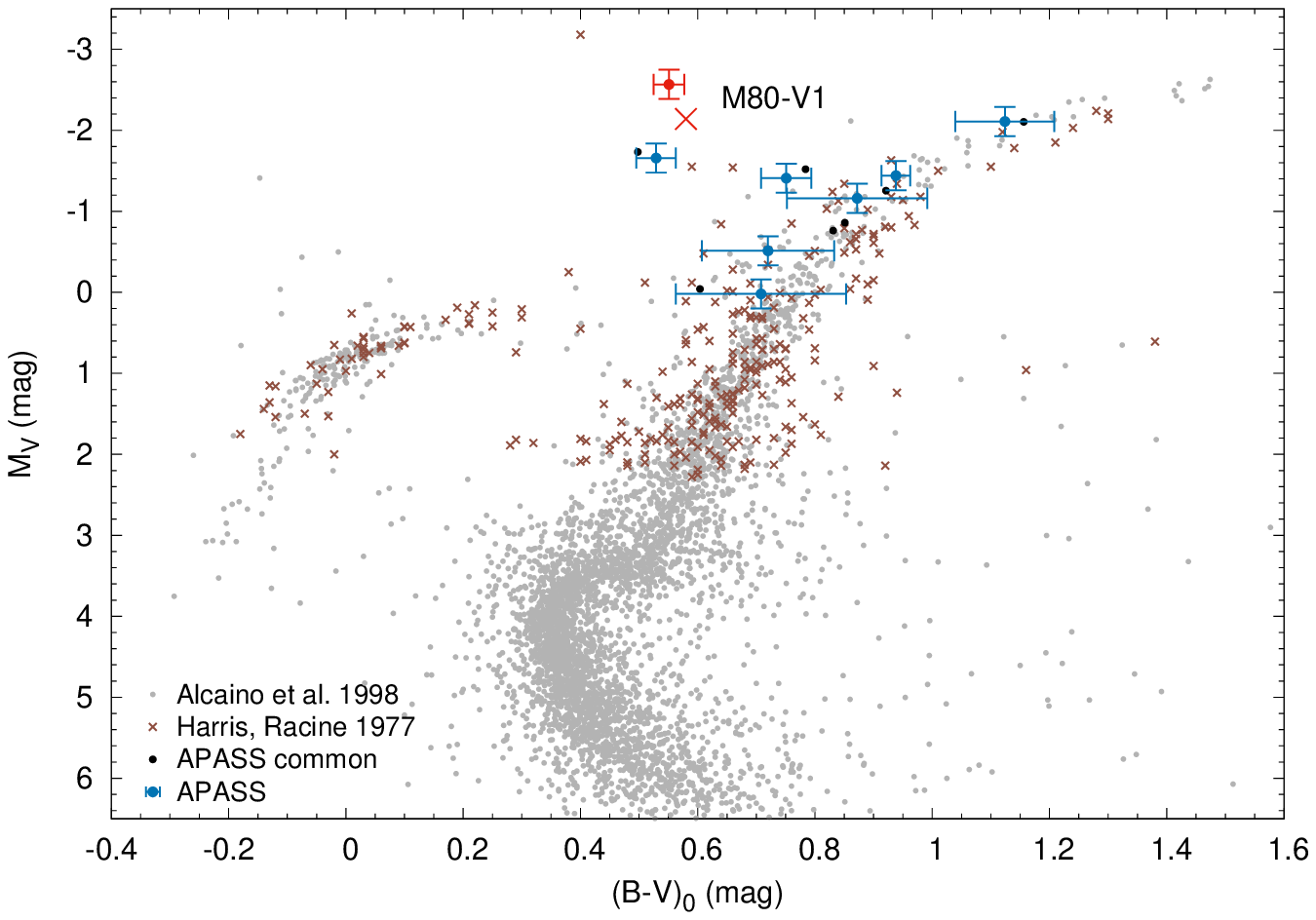}
\includegraphics[width=1.0\columnwidth]{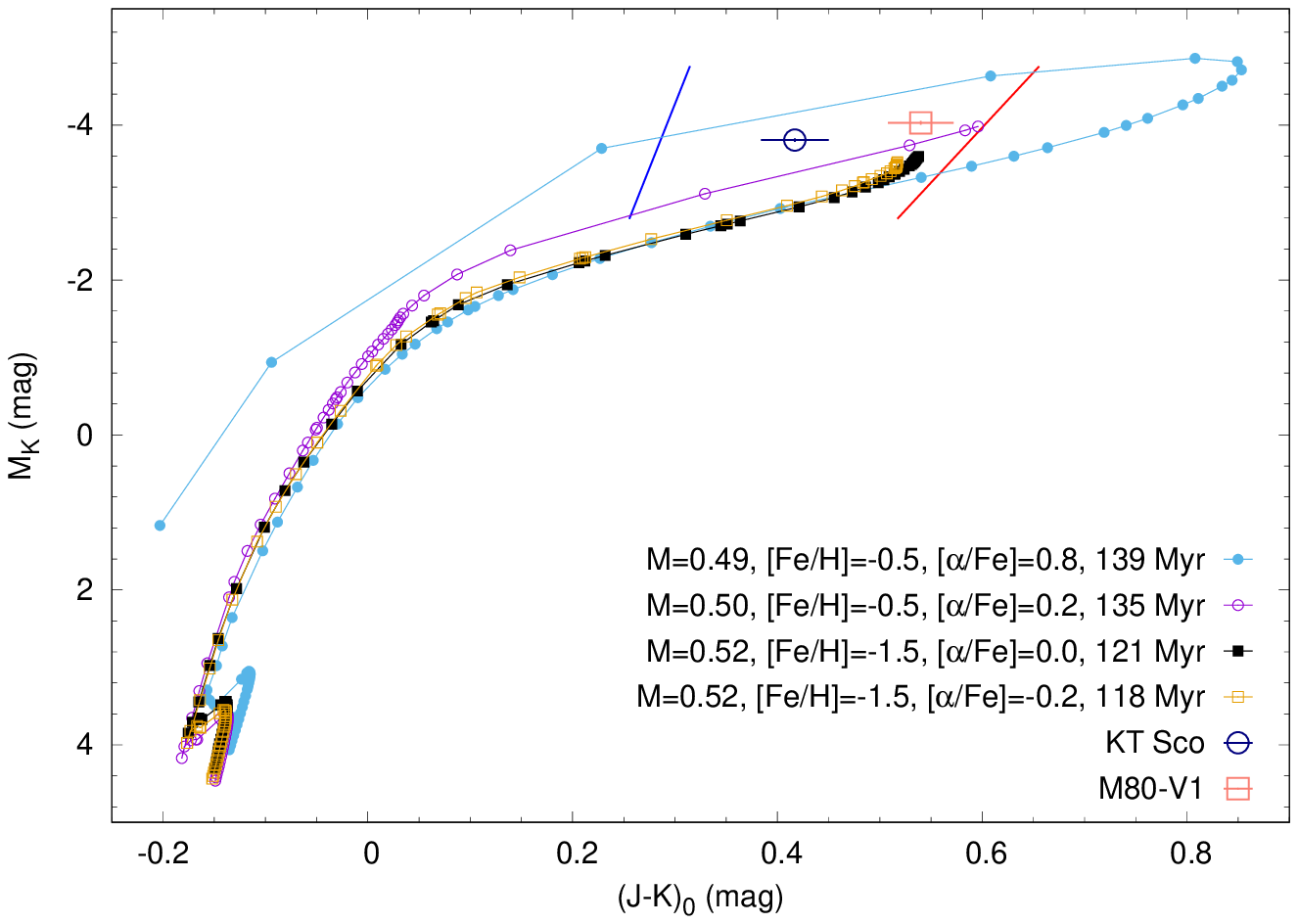}
\caption{Top: $M_V$--$(B-V)_§$ colour-magnitude diagram for M80. The red dot and red cross mark the position of M80-V1, based on the APASS data and the observations of \citet{kwee}. Larger black dots are stars from the catalogue of \citet{alcaino1998} that we identified in the APASS database (blue). Bottom: near-infrared $M_V$--$(B-V)_§$ colour-magnitude diagram for both stars. Blue and red lines mark the edges of the instability strip for RR Lyrae stars and Type II Cepheids. Dots and lines are evolutionary tracks from the Dartmouth horizontal branch stellar model library. We also indicated the post-ZHB ages where the models terminate. Instability strip borders (red and blue lines are from \citet{lemasle2015}).}  
\label{cmd1}
\end{figure}

To investigate the variability types and physical parameters of the stars, we decided to construct colour-magnitude diagrams (CMD) for them. Although M80-V1 is a member of the globular cluster, it lies at the outskirts of M80, and pulsates with a long period, so previous works either did not cover it or could not determine the average colour and brightness of the star \citep{harris1974,alcaino1998,piotto2002}. The work of \citet{harris1974} used the values determined by \citet{kwee}, and we found colour measurements for M80-V1 in the APASS (AAVSO Photometric All-Sky Survey) catalog too. We identified 7 stars that are present both in the APASS database and in the survey of \citet{alcaino1998} and have no close, bright neighbours, and thus can be used to crossmatch the two data sets. Unfortunately, the $B$-band data of APASS, and so the $(B-V)$ colour indices have large uncertainties in this brightness range. Nevertheless, we detected considerable differences between the two measurements. We shifted the latter data set to the APASS one (by adding 0.32 and 0.44 mag to the $B$ and $V$ measurements), because the older, photoelectric data of \citet{harris1974} agreed with the APASS brightnesses, and because \citet{alcaino1998} did not use standard stars from this field but relied on transformations obtained from other observations that could be the source of further uncertainties. We then transformed the apparent magnitudes and colours to absolute $V$ magnitudes and dereddened $(B-V)_0$ values. We used the distance modulus $(m-M)_V$ = 15.56 and interstellar extinction E$(B-V)$ = 0.181 to transform the data into absolute magnitudes and reddening-free colours, based on the catalog of \citet{harris1996} (2010 edition) and the \citet{schlafly-fink} extinction values from the IRSA Galactic Dust Reddening and Extinction tool\footnote{http://irsa.ipac.caltech.edu/applications/DUST/}.

In the final, $M_V$--$(B-V)_0$ diagram in the upper panel of Fig.~\ref{cmd1}, the position of the two measurements of M80-V1 are marked with a red cross and a red dot for the \citet{kwee} and APASS data, respectively. We determined the parameters of M80-V1 to be $M_V = -2.57 \pm 0.20$ mag, $(B-V)_0 = 0.551 \pm 0.027$ mag from the APASS observations, and $M_V = -2.14$ mag, $(B-V)_0 = 0.58$ from the older data (errors were not reported by \citet{kwee}).  These parameters put the star firmly into the middle of the high-luminosity part of the Type II Cepheid instability strip reported by \citet{rabidoux}. 

\begin{table}
\caption{Near-infrared absolute magnitudes, reddening-free colours, and respective uncertainties of M80-V1 and KT~Sco. Units are in magnitudes. }
\begin{center}
\begin{tabular}{ccccc}
\hline
\hline
Star& $M_K$ & $\Delta{M_K}$ & $(J-K_S)_0$ &  $\Delta(J-K_S)_0$ \\
\hline
M80-V1 & --4.028  & 0.023 & 0.540 & 0.032 \\
KT Sco & --3.806 & 0.024 & 0.417 & 0.033\\
\hline
\end{tabular}  
\end{center}
\label{nirtable}
\end{table}  

In order to determine the colour and luminosity properties of the field star KT Sco, and to compare the star to M80-V1, we examined the infrared measurements from 2MASS \citep{cutri2003,skrutskie2006} to construct a near-infrared CMD, following the procedure used by \citet{lemasle2015} for the Type II Cepheids DD~Vel and HQ~Car. The extinction correction was again gathered from IRSA for the 2MASS infrared bands \textit{J} and $K_S$ \citep{schlafly-fink}. Utilising the near-infrared period-luminosity relation derived by \citet{matsunaga}, we also calculated the $K_S$-band absolute magnitudes of the stars. The derived values are summarised in Table~\ref{nirtable}. The positions of the stars in the $M_K$--$(J-K_S)_0$ CMD were compared to some of the Dartmouth horizontal branch (HB) evolution stellar models \citep{dotter2008}. Most HB models crossed below the positions of the two stars, indicating that it is more likely that they have evolved into the instability strip at a later stage, rather than directly from the HB. We note that the evolution of W Vir-type variables is still under scrutiny, thus we refrained from a more detailed evolutionary study (for a recent summary, see \citealt{bono2016}, and references therein). Moreover, we also compared the two stars to the blue and red edges of the instability strip used by \citet{lemasle2015} that they also extrapolated slightly to higher luminosities. The CMD displayed in the lower panel of Fig.~\ref{cmd1} confirms that KT~Sco is a bona-fide W\,Vir-type star that is slightly hotter and fainter than M80-V1.

\section{Comparison with W V\lowercase{ir}}
Indications of cycle-to-cycle variations previously have been observed in W~Vir-type stars \citep{feast}. \citet{arp1955} noted that some Type II Cepheids with periods longer than 15 days may show alternating cycles. The OGLE survey found examples for changing or unstable W~Vir light curves, as well as PD in one case, but these have not been analysed in detail yet \citep{ogle2011}. 

Alternating cycles were observed in the prototype, W~Vir itself too \citep{abt1954}. \citet{templeton2007} collected extensive multicolour photometry of the star from Sonoita Research Observatory (SRO), and identified additional frequency peaks in the Fourier spectrum almost halfway between the main peak and its harmonics. However, as these peaks did not match the exact $(2n+1)/2\,f_0$ half-integer ratios, they concluded that those corresponded to the non-resonant first overtone and its combinations instead of PD. 

\begin{figure}
\includegraphics[width=1.0\columnwidth]{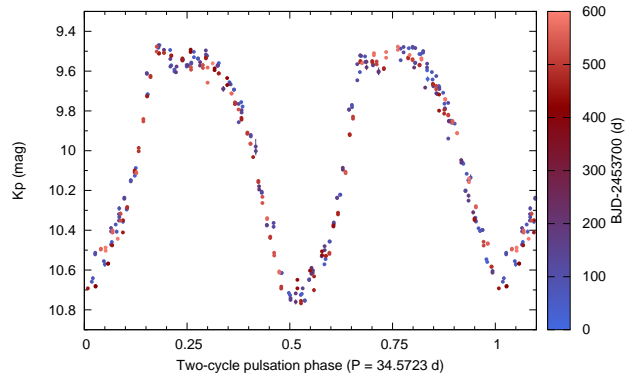}
\caption{Folded light curve of the \textit{V}-band SRO observations of W Vir, phased with twice the pulsation period. Note the differing depths of the odd and even minima caused by PD. Slight changes are suspected in the structure of the maxima that resemble the variations seen in the light curve of M80-V1 in Figure~\ref{ktsco_lc}. }
\label{wvir}
\end{figure}

Given the new findings about PD and related phenomena in classical variables, we decided to reanalyse the available data of W Vir. Our frequency fit to the combined $Hipparcos$, ASAS and SRO data revealed two half-integer frequencies at ratios $f/f_0$ = 0.50011 and 1.49996, respectively. We consider these frequencies as half-integer peaks as they are closer to the exact values than the frequency resolution of the data ($\Delta f = 0.00016$ d$^{-1}$, leading to an uncertainty of $\sim$0.004 in the frequency ratios). Although the light curves suggest some variation, the alternation of higher and lower minima apparently stayed in phase during the 6.45 years, or $\sim$136 pulsation cycles covered by the ASAS and SRO data (the sampling of the $Hipparcos$ data is too sparse to investigate). We note that in RV Tau stars, PD can lock into phase for even longer periods of time before the order of high and low minima turns over \citep{plachy2014}. When we analysed the SRO data that covers two seasons separately, we identified the half-integer peaks in all bands as well. We found the peaks again closer to the mathematical value than the frequency resolution of the data, therefore we can conclude that they may correspond to frequencies at the exact half-integer values that signal PD, and not just mode beating.  

Another argument against the overtone identification of the $1.5f_0$ is that the model calculations of \citet{model2} showed that the $P_0/P_1$ period ratio is only close to the 3:2 ratio for short-period BL Her stars, but it moves away from those values for longer periods. For models that are close to the 17.3 d period of W Vir, the $P_0/P_1$ ratio falls to the vicinity of the 2:1 resonance instead. Period doubling may arise from resonances with higher overtones in W Vir-type models \citep{model2,model_smolec}.

\section{Summary}
We analysed the first K2 observations of Type II Cepheids. These two stars, KT Sco and M80-V1, were observed in Campaign 2. We performed \textsc{Pyke} photometry with self-flat-field correction in the case of KT Sco, and standard aperture photometry for the globular cluster member M80-V1. The ultra-precise light curves and their Fourier transforms unveiled cycle-to-cycle variations in both stars: period doubling in M80-V1 and irregular changes of uncertain origin in KT Sco. We used previous ground-based observations to confirm the variability types of the stars and to study the long time-scale variations in their pulsation and found drastic phase changes in both stars, reaching $\sim$15 and $\sim$20 per cent for KT Sco and M80-V1, respectively. In the light of these findings we reanalysed the older, ground-based data of the prototype W Vir, and we concluded that period doubling can explain its light curve behaviour better than mode beating with a near-resonant first overtone. 

Our results suggest that dynamical instabilities play a role in the pulsation of W Vir stars confirming the predictions of the hydrodynamical calculations. We expect more findings of nonlinear phenomena in the future campaigns of the K2 mission that will observe hundreds of Type II Cepheids in the Galactic bulge.  



\section*{Acknowledgments}
We thank the comments of the anonymous referee that helped to improve the paper. We thank Arne Henden and Matthew Templeton for providing us with their multicolour photometry of W Vir, and James Nemec for his useful comments on the paper. This work has used K2 targets selected and proposed by the RR Lyrae and Cepheid Working Group of the \textit{Kepler} Asteroseismic Science Consortium (proposal numbers GO2039 and GO2041). Funding for the \textit{Kepler} and K2 missions is provided by the NASA Science Mission directorate. This project has been supported by the LP2012-31 and LP2014-17\linebreak Programs of the Hungarian Academy of Sciences, and by the NKFIH K-115709, PD-116175 and PD-121203 grants of the Hungarian National Research, Development and Innovation Office. L.M.\ was supported by the J\'anos Bolyai Research Scholarship of the Hungarian Academy of Sciences. M.I.J.\ acknowledges financial support from the Ministry of Education, Science and Technological Development of the Republic of Serbia through the project 176004. R.S. and P.A.M. are supported by the National Science Centre, Poland, through grants DEC-2012/05/B/ST9/03932 and DEC-2015/17/B/ST9/03421. The research leading to this results has received funding from the European Community's Seventh Framework Programme (FP7/2007-2013) under grant agreement no. 312844 (SPACEINN). The authors also acknowledge to ESA PECS Contract No. 4000110889/14/NL/NDe. Based on data from the OMC Archive at CAB (INTA-CSIC), pre-processed by ISDC. The CSS survey is funded by NASA under Grant No.\ NNG05GF22G issued through the Science Mission Directorate Near-Earth Objects Observations Program. This research was made possible through the use of the AAVSO Photometric All-Sky Survey (APASS), funded by the Robert Martin Ayers Sciences Fund. This research has made use of the NASA/ IPAC Infrared Science Archive, which is operated by the Jet Propulsion Laboratory, California Institute of Technology, under contract with NASA, and of the data products from the Two Micron All Sky Survey, which is a joint project of the University of Massachusetts and the Infrared Processing and Analysis Center/California Institute of Technology, funded by NASA and NSF.

\label{lastpage}

\end{document}